\documentclass[%preprint,twocolumn,
preprintnumbers,a4paper,superscriptaddress,aps,showpacs]{revtex4}
\usepackage{graphicx} % to include figure files
\begin{document}

\def\la{\langle}
\def\ra{\rangle}
\def\om{\omega}
\def\Om{\Omega}
\def\vep{\varepsilon}
\def\wh{\widehat}
\def\P0{\wh{\cal P}_0}
\def\dt{\delta t}
\newcommand{\beq}{\begin{equation}}
\newcommand{\eeq}{\end{equation}}
\newcommand{\beqa}{\begin{eqnarray}}
\newcommand{\eeqa}{\end{eqnarray}}
\newcommand{\intf}{\int_{-\infty}^\infty}
\newcommand{\into}{\int_0^\infty}
\title{A measurement-based approach to quantum arrival times}
\author{J. A. Damborenea}
\affiliation{Fisika Teorikoaren Saila, Euskal Herriko Unibertsitatea,
644 P.K., 48080 Bilbao, Spain}
\affiliation{Departamento de Qu\'\i{}mica-F\'\i{}sica, Universidad del
Pa\'\i{}s Vasco, Apdo. 644, 48080 Bilbao, Spain}
\author{I. L. Egusquiza}
\affiliation{Fisika Teorikoaren Saila, Euskal Herriko Unibertsitatea,
644 P.K., 48080 Bilbao, Spain}
\author{G.~C. Hegerfeldt}
\affiliation{Institut f\"ur Theoretische Physik, Universit\"at
G\"ottingen, Bunsenstr. 9, 37073 G\"ottingen, Germany}
\author{J. G. Muga}
\affiliation{Departamento de Qu\'\i{}mica-F\'\i{}sica, Universidad del
Pa\'\i{}s Vasco, Apdo. 644, 48080 Bilbao, Spain}

\begin{abstract}
For a quantum-mechanically spread-out particle we investigate a method
for determining its arrival time at a specific location. The procedure
is based on the emission of a first photon from a two-level system
moving into a laser-illuminated region. The resulting temporal
distribution is explicitly calculated for the one-dimensional case and
compared with axiomatically proposed expressions. As a main result we
show that by means of a deconvolution one obtains the well known
quantum mechanical probability flux of the particle at the location as
a limiting distribution.
\end{abstract}
\maketitle
\section{Introduction}

An important open problem in quantum theory is the question of how to
formulate the notion of ``arrival time'' of a particle, such as an
atom, at a given location, i.e. the time instant of its first
detection there. This is clearly a very physical question, but when
the extension and spreading of the wave packet is taken into account,
a satisfactory formulation is far from obvious. The problem of time in
quantum mechanics, both for time instants and time durations
such as dwell time,  has received a great deal of theoretical
attention recently \cite{ML00,MSE02}.  When the 
translational motion of the particle can be treated classically, a
full quantum analysis of arrival time is in fact not necessary. This
is the case for  fast particles, and therefore arrival times are
presently measured mostly by means of time-of-flight techniques, whose
analysis is carried out in terms of classical mechanics.  Problems,
though, arise for slow particles for which the finite extent of the
wavefunction and its spreading become relevant, such as for cooled
atoms dropping out of a trap. Then a quantum theoretical point of view
is needed.  It is 
therefore important to find out when the classical approximations
fail and to propose measurement procedures for arrival times in the
quantum case.  Since the theoretical definition of a quantum arrival
time is still subject to debate it is necessary to determine what
exactly such measurement procedures are measuring and to compare such
operational approaches with the existing, more abstract and axiomatic,
theories.

A simple one-dimensional example is the arrival time at $x=0$ of a
particle described by a wave packet moving to the right. The
probability of finding it still on the left side at time $t$ is
$\int_{-\infty}^0dx |\psi(x,t)|^2$. Then it would seem natural to
assume that the probability of arrival in $dt$ is given by the
decrease in $dt$ of this integral, i.e. that the arrival-time
probability density is its negative derivative. Since $\dot\rho +
dJ_{\psi}/dx = 0$, where $J_{\psi}$ is the usual quantum mechanical
probability current, one immediately finds that the arrival-time
probability density should be given by $J_{\psi}(0,t)$. For an
ensemble of classical particles the arrival time distribution is also
given by the flux, so everything seems to fit nicely. However, the
quantum flux for a  wavefunction formed entirely of positive-momentum
components may become negative at certain times 
 \cite{backflow}.  As a way out Leavens
\cite{Leavens} has proposed, following a Bohm trajectory analysis, to use its
normalized absolute value.
    
The difficulties to formulate a quantum arrival time concept were
posed most prominently by Allcock \cite{Allcock}, and there are
several attempts to overcome these \cite{ML00,BEM01d}. Kijowski
\cite{Kijowski74}, in  particular, obtained a 
time-of-arrival distribution for free motion from a set of axioms
modeled after the classical case; for a general investigation see Werner
\cite{Werner}.  The distribution of Ref. \cite{Kijowski74}  has been studied, 
compared to other approaches, and generalized by some of us for
systems subject to interaction potentials and to multi-particle
systems \cite{MLP98,MPL99,BSPME00,BEMS00,BEM01a,BEM01b}.

No procedure  how to measure the proposed 
distributions was 
given, nor is one known today. The gap between experiment and
axiomatically defined quantities has been commented on and considered
to be worrying by Wigner and others; for a review see
Ref. \cite{ML00}. In this vein, several ``toy models'' for
arrival-time measurements have been put forward by Aharonov,
Oppenheim, Reznik, Popescu and Unruh \cite{AOPRU98} (cf. also
\cite{ML00,BEM01c}), but these models do not incorporate the basic
irreversibility inherent in any measurement process.  Irreversibility
has been included by Halliwell \cite{Halliwell98b} in a model based on
a two-level detector in which the initial excited level decays due to
the presence of the particle.  However, the model remains somewhat
abstract since no connection is made with any specific measuring
system.

An experimentally very natural approach to determine the arrival time
of an atom is by quantum optical means. A region of space may be
illuminated by a laser and upon entering the region an atom will start
emitting photons. The first photon emission can be taken as a measure
of the arrival time of the atom in that region. This approach has been
proposed by three of us and Baute \cite{MBDE00}, and in a preliminary
study of the one-dimensional case a surprisingly good numerical
agreement with Kijowski's axiomatic distribution was found in some
special examples.

From a fundamental point of view, however, immediate objections may be
raised to this experimental, or operational, approach. First,
depending on the decay rate of the excited atomic level, the photon
emission will not be instantaneous but will take some time, thus
leading to a delay compared to some ``ideal'' arrival time of the
atom. Second, the laser takes some time to pump the atom from its
ground-state to an excited state, and therefore this also leads to a
delay. Conceivably, the second objection might be overcome by
progressively increasing the laser intensity, and the first objection
by considering shorter life times so that in a theoretical limit one
would arrive at an ``ideal'' quantum arrival time without the above
shortcomings. Attractive though this seems at first sight, it does not
work, as will be shown in this paper. The reason is a further
difficulty -- reflection. Although the laser couples only to the
internal degrees of the atom, it will be seen that there is a nonzero
probability for the atom to be reflected from the laser region without
ever emitting a photon.  Nevertheless, there is a way out of these
difficulties, with a surprising result. The idea is to ``subtract''
the delays from the first-photon probability density by means of a
deconvolution with an atom at rest. This results in a distribution
which, for shorter and shorter life time of the atomic level,
converges to an unexpected  distribution ~--~ namely to
$J_{\psi}$, the quantum mechanical probability flux. The 
probability distribution for the first photon is non-negative and the
emergence of possible small negative values is due to the
deconvolution procedure. 
This connection to $J_{\psi}$ opens a way, to our
knowledge for the first time, to measure the quantum mechanical
probability flux.  

For simplicity, this paper considers only the  one-dimensional
case. The probability density for the emission of the first photon from
a moving atom is calculated explicitly by means of the quantum jump approach 
\cite{Hegerfeldt93}. It is shown that  large laser intensities lead
to a large reflection probability. This in turn leads to a large
non-emission probability and a first-photon probability density not
normalized to 1. Then the problem of reflection versus time delay is
discussed. Reducing the laser intensity  leads to a pumping
delay. It is shown that trying to reduce the emission delay by shortening
the level lifetime leads in the limit to a free
wave packet in the ground-state with no emissions. The delays are then 
removed by a deconvolution,  and we discuss for which parameters
the resulting expression is close to its ``ideal'' limit $J_{\psi}$.
For more practical purposes  it is also
shown that for a certain domain of parameters, which include those
used in Ref. \cite{MBDE00}, the non-deconvoluted first-photon
probability density gives a good approximation to $J_\psi$  and to
Kijowski's axiomatic 
arrival time distribution. However, it is also pointed out that Kijowski's
distribution cannot be obtained in a simple direct way as an exact limit of
our operational approach.

\section{The probability density for the first photon}
The Hamiltonian of a two-level atom of mass $m$, interacting with the
quantized electromagnetic field ${\bf E}$ and a laser  with
(classical) field ${\bf E_L}$ is, in the 
usual dipole approximation and in the Schr\"odinger picture,
\begin{equation}
\label{2.1}
H = \frac{{\bf{\hat p}}^2}{2m} + H_A + H_F + e {{\bf D}} \cdot
\left({\bf{ E}}
% ( \hat{\bf x}, 0)
 + {\bf E}_L 
%(\hat{{\bf x}}, t)
\right)
\end{equation}
where
\begin{eqnarray}{\label{2.2}}
H_A &=& \frac{1}{2} \hbar \omega \left\{ |2\rangle\langle2| -
 |1\rangle\langle1| \right\} \\ \nonumber
H_F &=& \sum_{\bf{k} \lambda} \hbar \omega_{\bf{k}}
\hat{a}_{\bf{k}\lambda}^\dagger \hat{a}_{{\bf k}\lambda} \\\nonumber
{{\bf D}} &=& {\bf d}_{12} |1\rangle\langle2| + {\rm  h.c.}
\end{eqnarray}
with ${\bf d}_{12}$  the transition dipole moment between the
states $|1\rangle$ and $|2\rangle$. It is assumed that the laser illuminates a
half space, $x_1 \ge 0$, say.

Let an atomic state $|\Psi(t_0)\rangle$ be prepared at time $t_0$. By
means of the quantum jump approach \cite{Hegerfeldt93} the
atomic time development until the first photon detection is given by a
(non-hermitean) ``conditional'' Hamiltonian $H_{\rm c}$,%
\begin{equation}\label{2.3}
|\Psi(t)\rangle =
% e^{-iH_{A}(t-t_0)/\hbar} 
e^{-iH_{\rm c}(t-t_0)/\hbar} |\Psi(t_0)\rangle,
\end{equation}
 with the photon part traced away.
In the interaction picture with respect to the internal
Hamiltonian $H_A$  one has in the usual rotating wave approximation
\begin{equation}\label{2.4}
H_{\rm c} = {\bf{\hat p}}^2/2m + \frac{\hbar}{2} \Omega\, \Theta (\hat{x}_1)
\left\{ |2\rangle\langle1| e^{i{\bf k}_L\cdot {\hat{\bf x}}}+
{\rm h.c.}\right\}
 - \frac{i}{2} \hbar \gamma |2\rangle\langle2|~,
\end{equation}
where the Rabi frequency $\Omega \propto {\bf d} \cdot {\bf
E}^{(0)}_L$ plays the role of a laser-atom 
coupling constant (with ${\bf   E}^{(0)}_L$ the laser amplitude),
${\bf k}_L$ is the laser wave vector, and
where $\gamma$ is the Einstein 
coefficient of level 2, i.e. its decay rate or inverse life time. One
can show that Eq. (\ref{2.4}) includes the Doppler effect,
i.e. the laser driving depends on the atomic velocity through a
frequency shift. The probability, $N_t$, of no photon detection 
from $t_0$ up to time $t$ is given by \cite{Hegerfeldt93}
\begin{equation} \label{2.5}
N_t = || e^{-i H_{\rm c}(t-t_0)/\hbar} |\psi (t_0)\rangle||^2,
\end{equation}
and the probability density, $\Pi (t)$, for the first photon detection
by
\begin{equation} \label{2.6}
\Pi (t) = - \frac{dN_t}{dt}.
\end{equation}
%
%We will discuss further below to what extent $\Pi (t)$ can be
%considered to be an operationally defined arrival-time distribution.

For simplicity, we only consider the corresponding
one-dimensional problem, with the laser perpendicular to the atomic
motion, so that the Doppler effect plays no role. With $|1\rangle \equiv {1 \choose 0}$ and $|2\rangle \equiv {0 \choose 1}$
the conditional Hamiltonian becomes, in matrix form
\begin{equation}\label{2.7}
H_{\rm c} = \hat{p}^2/2m +\frac{\hbar}{2}\left({0\atop 0}{0 \atop
    -i\gamma} \right) + \frac{\hbar}{2}\,\Theta (\hat{x})
\left({0\atop \Omega}{\Omega  \atop 0} \right) .
\end{equation}
To obtain the time development of a general wave packet under $H_{\rm
  c}$ we first solve the eigenvalue equation
\begin{equation}\label{eigenvalue}
H_{\rm  c}{\bf \Phi} = E {\bf \Phi},~~~~~{\rm where}~~{\bf
  \Phi}(x)\equiv{\phi^{(1)}(x)\choose\phi^{(2)}(x)}. 
\end{equation}
Since for $x<0$ there is no laser one obtains for $\phi^{(i)}$ the
equations
\begin{equation}
\frac{\hat{p}^2}{2m} \phi^{(1)} = E \phi^{(1)},
~~~(\frac{\hat{p}^2}{2m}-i\hbar\gamma/2) \phi^{(2)} = E \phi^{(2)} .
\end{equation}
Hence $\phi^{(1)}$ is a combination of $e^{ikx}$ and $ e^{-ikx}$ with $k$
satisfying 
\begin{equation} \label{Ek}
E = \hbar^2k^2/2m \equiv E_k
\end{equation}
while $\phi^{(2)}$ is a combination of $e^{iqx}$ and $e^{-iqx}$ with $q$
satisfying 
\begin{equation} \label{Eq}
E + i\hbar\gamma/2 = \hbar^2q^2/2m ~. 
\end{equation}
We now look for eigenstates  of $H_{\rm c}$ which correspond to a
ground-state plane wave coming in from the left. Then $k$ as well as $E$ 
 must be real, by boundedness, and for $x \le 0$ the eigenstate is 
 of the form 
\begin{equation}\label{A15}
{\bf \Phi}_k (x) = \frac{1}{\sqrt{2\pi}} \left(  
{e^{ikx}+ R_1e^{-ikx} \atop R_2 e^{-iqx}}
 \right), \quad x\le0, \quad k>0, 
\end{equation}
with $  {\rm Im}\,q > 0$ for boundedness, while $R_1$ and $R_2$ are
reflection amplitudes yet to be determined. Note that although $E=E_k$
is real the complete wave functions will not be orthogonal, in
accordance with the non-hermiticity of $H_{\rm c}$.

To obtain the form of ${\bf \Phi}_k (x)$ for $x>0$ we denote by 
$|\lambda_+\rangle$ and 
$|\lambda_-\rangle$  the eigenstates of the
matrix $\frac{1}{2} \left( {0\atop \Omega}{\Omega\atop -i\gamma}
\right)$ corresponding to the eigenvalues $\lambda_\pm$. One easily finds
\begin{eqnarray}\label{A13}
\lambda_\pm &=& - \frac{i}{4}\gamma \pm \frac{i}{4}\sqrt{\gamma^2 - 4 \Omega^2}
\\ \label{A14}
|\lambda_\pm \rangle &=&  {1 \choose 2 \lambda_\pm/\Omega}~. 
\end{eqnarray}
We exclude, for the moment, the limiting case $\lambda_+ =
\lambda_-$. Note that $|\lambda_\pm \rangle$ are not orthogonal and
have not been normalized.
For $x \ge 0$, one can write ${\bf \Phi}_k$  as a superposition
of $|\lambda_\pm \rangle$ in the form
\begin{equation} \label{A17}
\sqrt{2\pi} {\bf \Phi}_k (x) = C_+ |\lambda_+ \rangle e^{ik_+x} +
C_- |\lambda_- \rangle e^{ik_-x}, \qquad x \ge 0.
\end{equation}
From the eigenvalue equation $H_{\rm  c}{\bf \Phi}_k = E_k {\bf \Phi}_k$,
together with $E_k = \hbar^2k^2/2m$, one obtains for $x \ge 0$
\begin{equation}\nonumber
k_\pm^2 = k^2 - 2m \lambda_\pm /\hbar = k^2 + im\gamma/2\hbar \mp
im  \sqrt{\gamma^2 - 4 \Omega^2}/2\hbar,
\end{equation}
with Im$\, k_\pm > 0$ for boundedness.
From the continuity of ${\bf \Phi}_k(x)$ at $x = 0$ one gets, with
$|1\rangle = {1 \choose 0}$  and $ |2\rangle = {0 \choose 1}$,
\begin{eqnarray}\nonumber
1 + R_1 &=& C_+ \langle 1|\lambda_+ \rangle + C_- \langle 1|\lambda_-
\rangle \\ \nonumber
R_2 &=& C_+ \langle2| \lambda_+\rangle + C_- \langle 2|\lambda_- \rangle.
\end{eqnarray}
Similar relations result from the continuity of ${\bf \Phi}_k'(x)$
at $x = 0$, yielding
\begin{eqnarray}\label{A18}
C_+ &=& -{2k(q+k_-)\lambda_-}/D,  \\ \nonumber
C_- &=&2k(q+k_+)\lambda_+/D, \\ \nonumber
R_2 &=& k(k_- -k_+)\Omega/D, \\ \nonumber
R_1 &=& [\lambda_+(q+k_+)(k-k_-)-\lambda_-(q+k_-)(k-k_+)]/D,
\end{eqnarray}
where the common denominator $D$ is given by
\begin{equation}\label{A19}
 D=(k+k_-)(q+k_+)\lambda_+
-(k+k_+)(q+k_-)\lambda_-\,.
\end{equation}
Thus Eq. (\ref{A17}) becomes, in components and for $x \ge 0$,
\begin{eqnarray}\label{A20}
\phi_k^{(1)}(x) &=& - \frac{2k}{\sqrt{2\pi}D}\left\{
(q + k_-)\lambda_- e^{ik_+x} - (q+k_+) \lambda_+ e^{ik_-x}
\right\}, \\ \nonumber
\phi_k^{(2)}(x) &=&  \frac{k\Omega}{\sqrt{2\pi}D}\left\{
(q + k_-) e^{ik_+x} - (q+k_+)  e^{ik_-x}
\right\}. \\ \nonumber
\end{eqnarray}
The case $\lambda_-=\lambda_+$ is obtained from this by taking
limits. For later purposes we also consider  increasingly  large $\gamma$, the 
other parameters kept fixed. This leads to  
\beqa
\label{first}
\lambda_+ &\approx& - i\Omega^2/2\gamma \to 0,
\\
\lambda_- &\approx& -  i \gamma/2 +
  i\Omega^2/2\gamma
 \to -i\gamma/2,
\\
k_+^2 &\approx& k^2 + im\Omega^2/\hbar\gamma,\quad \quad 
k_+\to k,
\\
k_-^2 &\approx&  k^2 + im \gamma/\hbar    \approx
im \gamma/\hbar,\\ \label{middle}
q &\approx& \left\{ im \gamma/\hbar\right\}^{1/2},
\\
C_-,&&\!\!\!\!\!\!\!\!\!R_{1},\, R_{2}\to 0,
\\
&&C_+\to 1.
\label{last}
\eeqa
In this case the state vector for $x>0$ becomes simply the plane wave 
with wave number $k$ in the ground state.  
This means that for increasing $\gamma$ there is less and less
reflection, but also less and less absorption, i.e. photon detection,
so that the laser has less and less effect on the atom.

At first sight the occurrence of reflections may seem surprising since
the laser only couples to the internal degrees of the atom and since
$H_{\rm c}$ only applies to the time development before the first photon
detection. Physically this can be understood from the coupling of the
atom to the quantized electromagnetic field. The laser changes the internal
state, this in turn changes the quantized electromagnetic field and
its momentum 
distribution. This in turn changes the momentum distribution of the
atomic motion. Mathematically the reason is of course the step
function in front of the matrix, similar as for a square-well potential. The
consequences of the nonzero reflection will be discussed further
below.

By decomposing an initial state as a superposition of eigenfunctions
one obtains its conditional time development. This is easy for an initial 
ground-state wave packet ${\psi (x,t_0)  \choose 0}$ coming in from
the far left side and with $t_0$ in the remote past.   Indeed, if
$\widetilde{\psi}(k)$ denotes the momentum amplitude the wave packet
would have as a freely moving packet at $t=0$, then
\begin{equation}\label{2.9}
{\bf \Psi}(x,t) = \int_0^\infty dk \,\widetilde{\psi}(k) \,{\bf \Phi}_k
(x)\,e^{-i \hbar k^2 t/2m}
\end{equation}
describes the conditional time development of a state which in the
remote past behaves like a wave packet in the ground-state coming in
from the left.

\section{The reflection problem and the no-detection probability}

From Eq. (\ref{2.5}) one obtains, with ${\bf \Psi} = {\psi^{(1)}
\choose \psi^{(2)}}$,
\begin{equation} \label{3.0}
N_t = \int dx \{ |\psi^{(1)} (x,t)|^2 + |\psi^{(2)} ({\bf x},t)|^2 \}
\end{equation}
and Eq. (\ref{2.6}) becomes
\begin{equation}\label{3.1}
\Pi (t) = \frac{i}{\hbar}
\langle\psi(t)|H_{\rm c} - H_{\rm c}^\dagger| \psi(t)\rangle.
\end{equation}
Since $H_{\rm c} - H_{\rm c}^\dagger = - i\gamma\hbar |2\rangle\langle2|$,
the first-photon probability density is given by
\begin{equation}\label{3.2}
\Pi (t) = \gamma \int_{-\infty}^\infty dx\, |\psi^{(2)}(x,t)|^2.
\end{equation}
The probability of no photon detection at all is, for $t_0$ in the
remote past, 
\begin{equation}\label{3.3}
1 - \int_{-\infty}^\infty dt'\, \Pi(t') = N_{t=\infty}.
\end{equation}
For physical reasons, only $\psi^{(1)}$ contributes to this, and only
for $x<0$. The latter follows from the fact that, for $x>0$, the
ground-state part will eventually be pumped by the laser to the excited
state. Since $t = \infty$ in
Eq. (\ref{3.3}), only the reflected part remains, and hence the
no-detection probability becomes
\begin{equation}\label{3.4}
N_{t=\infty} = \int_0^\infty dk\,|R_1(k)|^2 |\widetilde{\psi}(k)|^2.
\end{equation}
As a consequence, $\Pi (t)$ is in general not normalized to
$1$. Physically the probability for missing an atom increases with
$\Omega$, the strength of the laser driving. This is also seen
mathematically from the expression for $R_1$ in Eq. (\ref{A18}).
 An example is given in Section
\ref{negligible} (cf. Fig. \ref{jpi} further below) and a
practical approach to bypass this problem is also discussed there.

On the other hand, for $k\rightarrow \infty$ reflection becomes
negligible since then $R_i \rightarrow 0$. Hence for faster atoms
reflection does not pose a problem. This will also be exploited for
practical purposes in Section \ref{negligible}.
%\vspace{3cm}

%Do you want to insert numerical examples here???? 
%\vspace{3cm}

\section{Delays versus reflections and an idealized distribution} \label{delay} 

The approach to quantum arrival times by means of first-photon
detections contains a built-in ``delay'' since an excited atom will
not emit a photon immediately, due to  the finite decay rate
$\gamma$. This is in addition to the time the photon takes to reach
the detector; the photon travel time, however, can  easily be taken into
account, so it will not be considered here any further.

It seems natural to try to obtain an ideal arrival-time distribution
by considering faster and faster decay times, i.e. taking $\gamma
\rightarrow \infty$, at least theoretically. This will not work,
however. The reason is that for increasing $\gamma$, with all
other parameters kept fixed, the driving by the laser becomes less
efficient so that, in the limit $\gamma \rightarrow \infty$,
the wave packet remains unaffected, with no excitation and no
reflection, as can be seen from Eqs. (\ref{first}) - (\ref{last})
above. Moreover, if both $\gamma$ and $\Omega$ go to infinity with
$\gamma/\Omega$ kept fixed, then $R_1 \to -1$ and
everything is completely reflected without excitation \cite{repulsion}. 

\begin{figure}
%\epsfysize=7.6cm
%\centerline{\epsfbox{fmfig7.eps}}
\includegraphics[height=8cm]{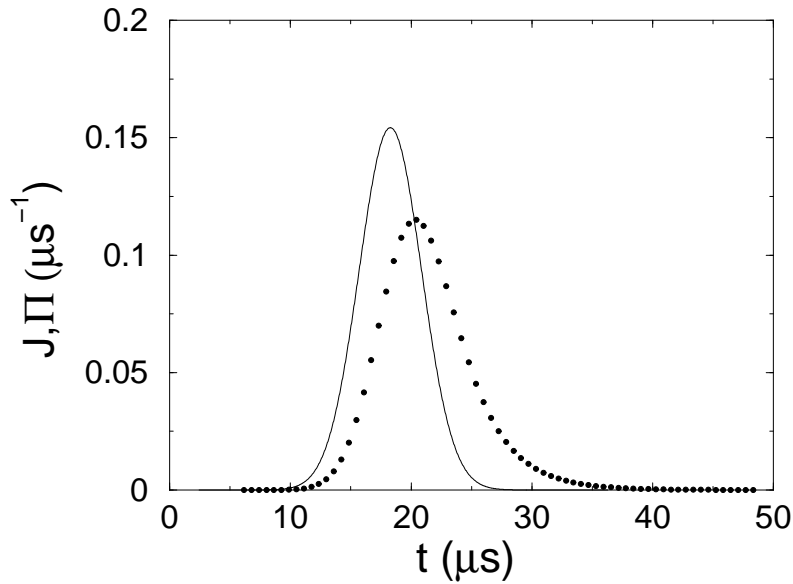}
\caption[]{Time-of-arrival distributions: Flux $J$ (solid line, here
  indistinguishable from Kijowski's $\Pi_K$)
 and $\Pi$ (first photon, dots). Note the delay in $\Pi$.
 The initial state is a minimum-uncertainty-product 
Gaussian  for the center-of-mass motion of a single Cesium atom in the
ground state with $\la v\ra=9.0297$ cm/s, 
$\la x\ra=-1.85$ $\mu$m, 
and $\Delta x=0.26$ $\mu$m.; $\Omega=0.0999\gamma$; all figures are 
for the transition $6^2P_{3/2}$ -- $6^2S_{1/2}$  of Cesium with $\gamma=33.3$
MHz.} 
\label{delnref}
\end{figure}

One might also be tempted to avoid reflection by choosing weak
driving, $\Omega/\gamma \ll  1$. This, however,
would cause a severe delay problem since the laser would
take more time to pump the atom to the excited state. Hence the
first-photon emission would also take more time. To see how relevant
detection delays are, we have compared $\Pi(t)$ with the flux $J_{\psi}$
and the axiomatic
probability distribution $\Pi_{\rm K}(t)$ of Kijowski for a Gaussian wave
packet. The result is given in Fig. \ref{delnref}.
Depending on the parameters,
the delay and reflection problem may be either very
relevant or negligible. A detailed analytic investigation of this
question is given in Section \ref{negligible}.

A way out of the conflicting problems of reflection (missed atom) and
 increasing delay times for weaker driving is the transition from the
``experimental'' $\Pi(t)$ to an idealized arrival-time distribution,
obtained as follows. A two-level atom at {\em rest}, when driven by a
resonant laser, 
has a definite probability density, $W(t)$, for the detection of the
first photon, given by \cite{2level}
\begin{equation}\label{A4}
W (t) = \frac{\gamma \Omega^2}{4|S|^2} e^{-\gamma t/2}|e^{St/2} -
e^{-St/2}|^2 \Theta (t),
\end{equation} 
with
\begin{equation}\nonumber
S = \frac{1}{2} (\gamma^2 - 4 \Omega^2)^{1/2}.
\end{equation}\nonumber
 Intuitively,
the delay-time mechanism for a moving atom ought to
be similar to that for an atom at rest. Should it then not be
possible to somehow compensate the  delay in $\Pi(t)$ by that of the
atom at rest and thus arrive, in some limit, at a
delay-free ideal distribution? To achieve this, we assume
the (experimental) arrival-time distribution $\Pi(t)$ to be the
convolution of a hypothetical ideal distribution, $\Pi_{\rm id}$, with the
distribution $W(t)$ for an atom at rest,
\begin{equation} \label{4.1}
\Pi = \Pi_{\rm id}*W.
\end{equation}
The delay in $\Pi$ is then mainly due to that contained in $W$.

The hypothetical 
ideal distribution $\Pi_{\rm id}$ is obtained by a deconvolution via
Fourier transform from $\widetilde{\Pi}_{\rm id} =
\widetilde{\Pi}/\widetilde{W}$ where ${\widetilde \Pi}(\nu) = \int dt
e^{-i\nu t} \Pi (t)$ etc. From Eqs. (\ref{3.2}) and (\ref{2.9}) one finds
\begin{equation}\label{A3}
{\widetilde \Pi}(\nu)
= \gamma \int dx  \int dk\int dk'\, \overline {{\widetilde
\psi} (k)} 
{\widetilde\psi} (k') \overline{\phi_k^{(2)}(x)} \phi^{(2)}_{k'}(x)  2
\pi \delta 
\left( \nu - \frac{\hbar}{2m} (k^2 - k'^2)    \right)
\end{equation}
and from Eq. (\ref{A4})
\begin{equation}
\label{A5}
\widetilde{W} (\nu) = \frac{\Omega^2\gamma/2}{\left\{ 
\left(  i \nu +\gamma/2  \right)^2  - S^2\right\} 
\left\{ i\nu + \gamma/2 \right\}},
\end{equation} 
and thus
\begin{equation} \label{A6}
\frac{1}{\widetilde{W}(\nu)} = 1 + C_1 i\nu + C_2(i\nu)^2 + C_3(i\nu)^3,
\end{equation}
where
\begin{equation}\label{A7}
C_1 = \left( \frac{\gamma}{\Omega^2} + \frac{2}{\gamma} \right),\quad
C_2 = \frac{3}{\Omega^2}~, \quad C_3 = \frac{2}{\gamma \Omega^2}~.
\end{equation}
In the time domain this gives
\begin{equation}\label{4.1a}
\Pi_{\rm id}(t) = \Pi (t) + C_1 \Pi'(t) + C_2 \Pi''(t) + C_3 \Pi'''(t)~.
\end{equation}

 In the limit of no reflection, for $\gamma \rightarrow \infty$, the  delay
problem pointed out above for $\Pi$ should be absent for $\Pi_{\rm
  id}$. In fact, in this limit one sees from Eqs. (\ref{first}) -
(\ref{middle}) and from Eq. (\ref{A6}) that in $1/{\widetilde W}$ only $i \nu
\gamma/\Omega^2$ remains relevant for 
large $\gamma$. Furthermore,  $R_2 \to 0$ for
$\gamma \rightarrow \infty$, and since $e^{-iqx} \rightarrow 0$ in
Eq. (\ref{A3}) the integral over $x 
\le 0$ goes to zero. Similarly the terms in $\phi^{(2)}_k(x)$ and
$\phi^{(2)}_{k'}(x)$ containing $e^{-i \bar{k}_-x}$ and $e^{ik'_-x}$ drop
out. For $\gamma \rightarrow \infty$ one has, from Eqs. (\ref{first})
and (\ref{last}), $C_+
\rightarrow 1$ and the integration of $\exp \{- i \bar{k}_+ x + i k'_+
x \}$ over $x \ge 0$ gives $i/(k'_+ - \bar{k}_+)$. Hence
\begin{equation}\label{A10}
\frac{\widetilde{\Pi} (\nu)}{\widetilde{W}(\nu)} \rightarrow \frac{4\gamma}{2
  \pi} \int dkdk' \overline{{\widetilde \psi}(k)} \widetilde{\psi}(k') 
\frac{\bar{\lambda}_+ \lambda_+}{\Omega^2} \frac{i}{k_+' - \bar{k}_+}
\frac{i\nu \gamma}{\Omega^2}  2 \pi \delta 
\left( \nu - \frac{\hbar}{2m} (k^2 - k'^2)
\right).
\end{equation}
By the $\delta$ function, $\nu = \hbar (k - k')(k + k')/2m$. Hence in
the limit $\gamma \rightarrow \infty$ one obtains, by
Eqs. (\ref{first}) - (\ref{middle}),
\begin{equation}\label{A10a} 
\frac{\widetilde{\Pi}(\nu)}{\widetilde {W}(\nu)} \rightarrow
\frac{1}{2 \pi} \frac{\hbar}{2m} \int dkdk' \overline{\widetilde{\psi}
  (k)}{\widetilde \psi} (k') (k+k') 2 
\pi \delta \left( \nu - \frac{\hbar}{2m} (k^2-k'^2)   \right).
\end{equation}
Going to the time domain one finally obtains
\begin{eqnarray}\label{4.2}
\Pi_{\rm id} (t) &\rightarrow& \frac{1}{2\pi} \frac{\hbar}{m} \int dkdk'
\overline{\widetilde{\psi}(k)} e^{i \hbar k^2t/2m}\frac{k +
  k'}{2}{\widetilde \psi} (k') e^{-i \hbar k'^2 t/2m} 
\\ \nonumber
&=& \frac{\hbar}{2mi} \left\{ \overline{\psi (0,t)}  \psi'(0,t) -
\overline{\psi' (0,t)} \psi (0,t)  \right\},
\end{eqnarray}
which is the flux $J_\psi$ for the {\em free}
wavefunction $\psi 
(x,t)$ at $x = 0$, i.e. without laser \cite{footnote1}.

This is an extremely
interesting result since $J_\psi$ is a natural candidate for the
arrival-time distribution, as pointed out in the Introduction. 
%Its only drawback is that for particular wavefunctions it may have small
%negative parts. 
We note that $J_\psi$ is normalized to 1 for a particle which has only positive
momentum components. This is seen for example from Eq. (\ref{A10a})
for $\nu=0$. 

\begin{figure}
%\epsfysize=7.6cm
%\centerline{\epsfbox{....eps}}
\includegraphics[height=8cm]{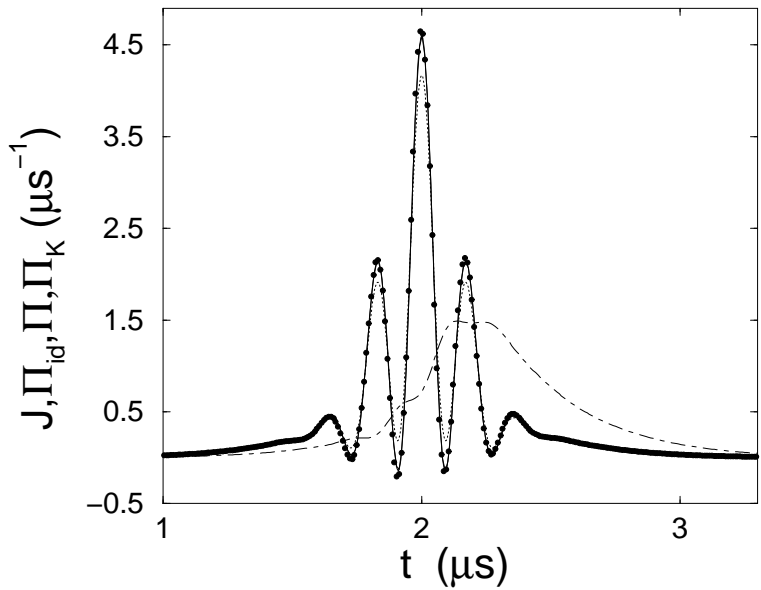}
\caption[]{Excellent agreement between $\Pi_{\rm id}$ (filled circles)
  and $J$ (solid line); deviations from $\Pi_K$ (dotted line) and  
$\Pi$ (dot-dashed line). 
The initial wave packet is 
a coherent combination $\psi=2^{-1/2}(\psi_1+\psi_2)$
of two Gaussian states for the center-of-mass motion of a single Cesium atom
that become separately minimal uncertainty packets (with
$\Delta x_1=\Delta x_2=0.021\,\mu$m,
and average velocities 
$\la v\ra_1=18.96$ cm/s, 
$\la v \ra_2 =5.42$ cm/s) at $x=0$ and $t=2\,\,\mu$s;
$\Omega=0.37 \gamma$.
}
\label{deconv}
\end{figure}

The limit in Eq. (\ref{4.2}) means that $\Pi_{\rm id}$ can be approximated
by $J_\psi$ for sufficiently large $\gamma$. Physically, it is
important to determine the parameter ranges for which this approximation is
a good one.  For this to be valid a simple sufficient condition
on the parameters 
can be derived as follows. First one considers the case
$\Omega^2/\gamma^2 \ll 1$ (weak driving) and makes the corresponding
approximations in $\Pi_{\rm id}(t)$. Denoting by $\Delta E$ a
measure of the magnitude of
\begin{equation}\label{A12}
\frac{\hbar^2}{2m} |k^2 - k'^2|\, ,
\end{equation}
as determined by $\widetilde{\psi}(k)$ and $\widetilde{\psi}(k')$, one then
considers the case $\gamma \gg \Delta E/\hbar$ and $\Delta E/\hbar \gg
\Omega^2/\gamma$. Then one obtains that $\Pi_{\rm id}$ is close to
$J_{\psi}$. The inequalities can be written in the form 
\begin{equation}\label{4.3}
\Omega^2/\gamma \ll \Delta E/\hbar \ll \gamma
\end{equation}
 Thus $\Pi_{\rm id}$ can 
be replaced by $J_\psi$ if these inequalities are satisfied. Even
outside this parameter 
range, $J_\psi$ may be an excellent approximation to $\Pi_{\rm id}$,
as shown in Fig. \ref{deconv}.
The convergence of $\Pi_{\rm id}$ to $J_\psi$ shows that also
$\Pi_{\rm id}$ may contain small negative values. Since this occurs neither
for $\Pi$ nor for $W$, the intuitive ansatz of Eq. (\ref{4.1}) cannot
always be fulfilled with a strictly positive distribution. The reason
for this clearly is that the ansatz of a convolution in Eq. (\ref{4.1})
is too simple and ought to be replaced by something more
sophisticated. On the other other hand, this result gives a handle at
the quantum mechanical probability flux and indicates a method how to
measure it. 

 The above deconvolution procedure which recovers the quantum
 mechanical particle flux essentially works because the
 weak-excitation limit taken allows a clear separation of (1) the time
 dependence associated with the 
 motion of the wavepacket, and (2) the time dependence associated with
 the internal degrees of freedom (excitation, Rabi oscillation, and
 decay). It seems reasonable that this might be done. However, the
 weak-excitation limit implies that the waiting times for the 
 first scattered photon are of the order $\gamma/\Omega^2$ and therefore
 very long. Hence the number to be measured  is
 essentially to be obtained from the subtraction of two very large
 numbers. Experimentally this is a difficult thing to do with high
 accuracy and requires small measurement errors. For practical
 purposes it is therefore important to know when  delays and reflections 
can be safely neglected, since then the transition to $\Pi_{\rm id}$
by deconvolution is not necessary. This is investigated quantitatively
in the next section. 
\section{Parameter ranges with negligible delay and
reflection.} \label{negligible}

In the case of weak driving, $\Omega\ll \gamma$, the excited state
population is negligible  compared to that of the ground state. Hence the  
reflection coefficient for the ground state is the only one that 
matters in this case,  
\beqa\label{r112}
|R_{1}(k)|^2\approx \frac{1}{64}(\Omega / \gamma)^2(\hbar \Omega/E)^2,   
\eeqa
This is small if $E\gtrsim \hbar\Omega$. For strong driving,
$\Omega\gg \gamma$,  the reflection coefficients 
take a simple form when the energy E of the plane wave satisfies
\beqa\label{r112a}
    E \gg\hbar\Omega.
\eeqa
Then both states are populated roughly equally for  
$x>0$, and with Eq. (\ref{r112a}) the reflection coefficients become 
\beq\label{r111} 
|R_{1}(k)|^2\approx\frac{1}{32^2}(\hbar \Omega/E)^4,\;\;\;
|R_{2}(k)|^2\approx \frac{1}{64}(\hbar \Omega/E)^2. 
\eeq
Both coefficients are small if Eq. (\ref{r112a}) holds.

To quantify the detection delay,  
let us define $\tau_d$ as the difference between the average 
time of the  first photon emission, 
$\la t\ra_{\Pi}=\intf dt\,t\,\Pi (t)$,
and $\la t\ra_J=\intf dt\,t\, J_{\psi}$ \cite{footnote2}; 
we note that
the ``average arrival time'' at $x=0$ evaluated with the flux at $x=0$, 
$\la t\ra_J$ coincides with the average of
Kijowski's distribution \cite{Kijowski74,ML00}.
For negligible reflection, ${\bf \Psi}(x,t)$ is, for $x<0$, nearly the
same as the free wavefunction, and therefore, by a partial
integration,
\beq\label{taud}
\tau_d=\la t\ra_{\Pi} - \la t\ra_J = \int_{-\infty}^{\infty} dt\,N_t^+~~~~{\rm (for~
  negligible~ reflection)},     
\eeq
where $N_t^+=\into (|\psi^{(1)}(x,t)|^2+ |\psi^{(2)}(x,t)|^2)\,dx$.  
Hence, in the case of negligible reflection the delay is associated
with the amount of penetration of the wave into the laser region.
For weak and strong driving a straightforward calculation yields the
simple results  
\beqa\label{gaom2}
\tau_d&\approx&\gamma/\Omega^2\;\;{\rm(weak\; driving}).
\\
\tau_d&\approx&2/ \gamma\;\;\,{\rm (strong\; driving),} 
\eeqa
It is interesting to note, and physically very reasonable, that this
coincides with the average time between two photon emissions for a
two-level atom at rest, driven by a resonant laser, as easily seen from
Eq. (\ref{A4}). If $\Delta t$ denotes the width of $\Pi(t)$, then one
needs  $\tau_d\ll \Delta t $ for  the delay to be negligible.

\begin{figure}
%\epsfysize=7.6cm
%\centerline{\epsfbox{fmfig9.eps}}
\includegraphics[height=6.8cm]{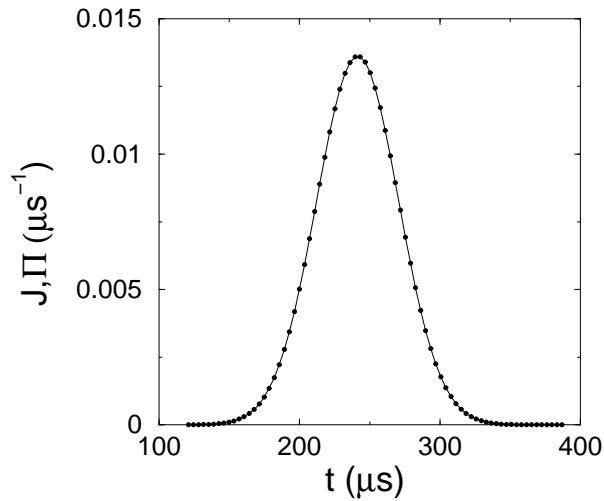}
\caption[]{Negligible delay and reflection with Eq. (\ref{strong}):
$\Pi$ (first photon, dots) and $J_\psi$ (solid line, here
indistinguishable from $\Pi_K$)  for initial Gaussian parameters 
$\la v\ra=90.30$ cm/s, $\la x\ra=-218.02$ $\mu$m, 
$\Delta x=26.46$ $\mu$m; $\Omega=5 \gamma$.}
\label{ndelnref}
\end{figure}

If one denotes by $\widetilde{E}$  the average energy for  
wave packets that are sharply peaked in energy or, 
more generally, as the smallest significant energy component, then
the above conditions for negligible reflection
and negligible delay  can be summarized as
\beqa 
\hbar/{\widetilde{E}} \lesssim \gamma/\Omega^2&\ll&\Delta t
%>\hbar/{\widetilde{E}}
\;\;\;\;\;\;\;
\;\;\;  ~~~~~~~~~ ({\rm weak\,\, driving}) 
\\ \label{ineq1}
\hbar/\widetilde{E}\ll \Omega^{-1}&\ll& 2/\gamma\ll \Delta t 
\;\;\; ~~~~   ({\rm strong\,\,
  driving}).\label{strong}
\eeqa
Fig. \ref{ndelnref} shows a striking example for which these conditions are fulfilled. 
Here the first-photon distribution $\Pi$ is indistinguishable from
$J_{\psi}$ and from Kijowski's distribution. 

\begin{figure}
%\epsfysize=7.6cm
%\centerline{\epsfbox{....eps}}
\includegraphics[height=8cm]{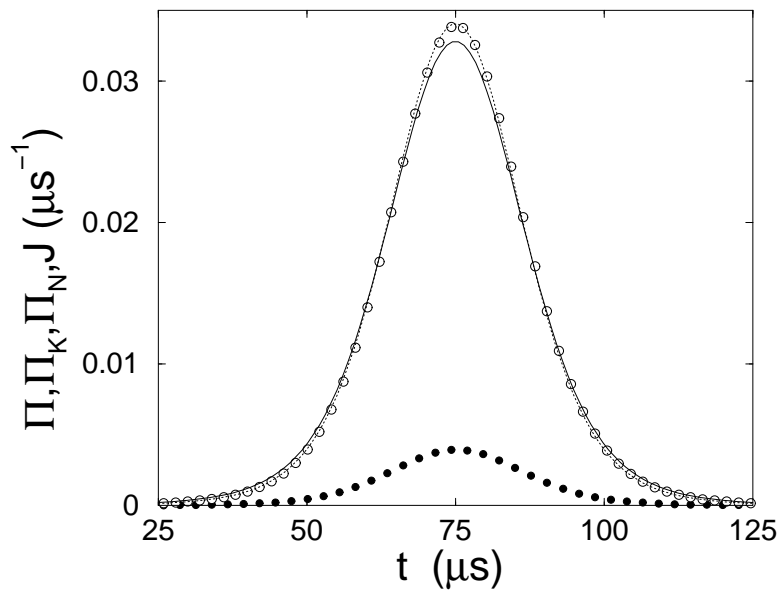}
\caption[]{Improvement by normalization:  $\Pi_N$~(white circles)
  right on top of $J_\psi$~(dashed line); also shown  
$\Pi_K$~(solid line) and $\Pi$~(filled circles).
  The initial Gaussian wave packet is chosen to become 
minimal when its center arrives at $x=0$ (in the absence of the
laser) to enhance the difference between $\Pi_K$ and $J_\psi$;
 $\la v\ra =0.9$ cm/s, 
$\Delta x=0.106$ $\mu$m, $\Omega=3\gamma$.}
\label{jpi}
\end{figure}

For strong driving the delay due to the time needed for pumping the
atom to an excited level is small, but on the other hand there is a
large probability to miss out the atom altogether if the condition 
$E\gg\hbar\Omega$ is not fulfilled. As a consequence, the
first-photon probability density $\Pi(t)$ will not be normalized to
1. Let us then define a
normalized distribution $\Pi_N(t)= \Pi(t)/\int dt'\,\Pi(t')$.
For practical purposes
this can be amazingly efficient in some parameter domains as  Fig. \ref{jpi}
shows. There, $\Pi$ is far from being normalized to 1, 
but $\Pi_N$ and $J_\psi$, which in this example has no negative parts,
coincide beautifully. Interestingly, one notices
a small difference to Kijowski's distribution. It might also be
possible to use the general normalization procedure  in terms of
operators proposed in Ref. \cite{BF01}.
\section{Conclusions}\label{concl}
In this paper, we have investigated a proposal to determine
arrival times of quantum mechanical particles. The proposal is based
on the intuitive idea to
illuminate the arrival region by a  laser and  to consider a
traveling two-level atom. The time of the first emitted photon is then
taken as a measure for the arrival time. By repeating the experiment one 
obtains a probability density, $\Pi (t)$, for the time of the first photon.
We have discussed for the one-dimensional case in what way $\Pi (t)$ can
be regarded as an atomic arrival-time distribution. Restrictions
arise from reflections and delays.
Reflections 
 originate from the interaction with the laser and
delays from the time needed for the  pumping and the ensuing
photon emission. The natural idea that an ideal or an axiomatically
proposed distribution might be obtained from 
$\Pi (t)$ in the limit of a very strong or very weak laser and very large
Einstein decay coefficient of the excited level has turned out not to
be true. However, and
this is a main theoretical result of the paper, one can subtract the delay by a
deconvolution with the first-photon probability density for an atom at
rest and then,  surprisingly, for larger and larger Einstein
coefficient  one obtains the quantum mechanical probability current $J$ as the
limit distribution. This quantity $J$ has previously been considered on
axiomatic grounds as a candidate for the arrival time
distribution, and the connection of $\Pi(t)$ with $J$ indicates, to our
knowledge for the first time, a way to
measure the quantum mechanical probability flux. 
We have also determined parameter domains for which the
deconvoluted expression is already sufficiently close to $J$.
Although the non-deconvoluted $\Pi (t)$ is not the same as $J$
and the axiomatically
proposed distribution of Kijowski, it can, for
experimental purposes, approach the latter two sufficiently closely. Parameter
domains for which this holds have been explicitly determined; this is
another main result of the paper, more of a practical nature. 

\begin{acknowledgments}
This work has been supported
by Ministerio de Ciencia y Tecnolog\'\i a (BFM2000-0816-C03-03), 
UPV-EHU (00039.310-13507/2001),
and the Basque Government (PI-1999-28).
\end{acknowledgments}

%\newpage


\begin{thebibliography}{10}
\bibitem{repulsion} More precisely,
if $\Omega$ and $\gamma$ are of the same order of magnitude and
$\gamma \gg \frac{\hbar k}{m} k$ or, alternatively, if $\Omega /\gamma
\ll 1$ and $ \Omega^2/\gamma \gg \frac{\hbar k}{m} k$, then $R_1
\approx -1$ . Physically, both conditions mean that the distance 
traveled by the atom in the time it takes for a photon to be scattered is much
less than the de Broglie wave length.
\bibitem{footnote1} For wavefunctions with positive momenta, Kijowski's
distribution is obtained by replacing $\frac{k + k'}{2}$ with
$\sqrt{kk'}$ in Eq. (\ref{4.2}). The occurrence of a square root of
$k$ shows that this distribution cannot be obtained from $\Pi$ or
$\Pi_{\rm id}$ in a simple direct way as  a limiting case since the latter
contain only integer powers of $k$.
\end{thebibliography}
\end{document}